\documentclass[aps,prl,twocolumn,showpacs,superscriptaddress,groupedaddress]{revtex4}
\usepackage{graphicx}
\usepackage{dcolumn} 
\usepackage{bm}    
\usepackage{amssymb}  
\usepackage{amsmath}

\begin{document}

\title{Simultaneous Acceleration of Protons and Electrons\\ 
at Nonrelativistic Quasiparallel Collisionless Shocks}

\author{Jaehong Park, Damiano Caprioli, \& Anatoly Spitkovsky}

\affiliation{Department of Astrophysical Sciences, Princeton University, Princeton, NJ 08544, USA}

\date{\today}

\begin{abstract}

We study diffusive shock acceleration (DSA) of  protons and electrons at nonrelativistic, high Mach number, quasiparallel, collisionless shocks by means of self-consistent 1D particle-in-cell simulations. 
For the first time, both species are found to develop power-law distributions with the universal spectral index $-4$ in momentum space, in agreement with the prediction of DSA. 
We find that scattering of both protons and electrons is mediated by right-handed circularly polarized waves excited by the current of energetic protons via non-resonant hybrid (Bell) instability.  
Protons are injected into DSA after a few gyro-cycles of shock drift acceleration (SDA), while electrons are first pre-heated via SDA, then energized via a hybrid acceleration process that involves both SDA and Fermi-like acceleration mediated by Bell waves, before eventual injection into DSA. 
Using the simulations we can measure the electron/proton ratio in accelerated particles, which is of paramount importance for explaining the cosmic ray fluxes measured at Earth and the multi-wavelength emission of astrophysical objects such as supernova remnants, radio supernovae, and galaxy clusters.  We find the normalization of electron power-law is $\lesssim 10^{-2}$  that of the protons for strong nonrelativistic shocks. 
\end{abstract}

\maketitle

 \emph{Introduction.---} Diffusive shock acceleration (DSA) \citep[e.g.,][]{bell1978,blandford1978} at supernova remnant (SNR) shocks is widely regarded as the mechanism responsible for the acceleration of Galactic cosmic rays (CRs) up to $E\sim10^{17}$ eV.
The presence of multi-TeV electrons and protons is also revealed through copious broadband non-thermal emission from SNRs.
In DSA, particles gain energy by repeatedly scattering across the shock, increasing their energy as if being squeezed between two converging walls.
Their final momentum distribution $f(p)$ is a universal power-law whose spectral index depends on the shock hydrodynamics:
for strong shocks (and for gas adiabatic index $5/3$), the compression ratio $r\to 4$, and $f(p)\propto p^{-3r/(r-1)}\propto p^{-4}$.

One of the most important questions in CR physics is how particles are extracted from the thermal pool to be accelerated to relativistic energies, i.e., how particles are injected into DSA. 
Recent hybrid simulations of high Mach number shocks \citep{caprioli2014a,caprioli2014b,caprioli2014c,caprioli2014d} showed that in shocks propagating in the direction quasiparallel to the background magnetic field, protons are injected after having been specularly reflected at the shock potential barrier and energized via shock drift acceleration (SDA) up to injection momentum $p_\text{inj}\approx 2.5m_pv_\text{sh}$, where $m_p$ is the proton mass and $v_\text{sh}$ is the shock velocity. After this, protons begin to diffuse around the shock. 
Achieving such a $p_\text{inj}$ is much more difficult for electrons, as their initial momentum is a factor of $m_p/m_e$ smaller; 
therefore, it is natural to expect a preferential injection of protons over electrons.
Indeed, the electron/proton ratio in accelerated particles, $K_\text{ep}$, is consistently inferred to be much smaller than one.  Direct detection of CRs at Earth results in $K_\text{ep}\approx 0.01$ around 10GeV, where solar modulation and electron radiative losses are negligible \citep[e.g.,][]{pamela}, while multi-wavelength observations of young SNRs suggest $K_\text{ep}\approx 10^{-3}$ or less \citep[see, e.g.,][]{V+05,morlino2012}.

Understanding electron injection requires a self-consistent calculation of the electromagnetic shock structure along with the dynamics of both protons and electrons, which can be achieved only with kinetic simulations that can capture the nonlinear interplay between energetic particles and self-generated fields over a broad range of time and length scales.
Although several simulation studies have already reported non-thermal particle acceleration in diverse collisionless shock environments \citep[e.g.,][]{riquelme2011,kato2014,guo2014a,guo2014b}, first principles simulations have never shown simultaneous DSA of both protons and electrons in non-relativistic collisionless shocks; consequently, $K_\text{ep}$ has never been measured in ab-initio simulations.

In this Letter, we report on the use of large particle-in-cell (PIC) simulations to study both proton and electron acceleration in nonrelativistic, quasiparallel, strong shocks relevant for young SNRs, demonstrating for the first time the formation of universal $\propto p^{-4}$ DSA spectra for both species.
We also characterize electron injection, outlining the crucial role of the nonresonant modes excited by the proton-driven Bell instability \cite{bell2004} in shaping the shock dynamics and in regulating particle scattering, and measure the electron/proton ratio in energetic particles.
Finally, we discuss the application of our findings to the phenomenology of radio-SNe and young SNRs.

\emph{PIC Simulations.---} We use the parallel electromagnetic PIC code TRISTAN-MP \citep{buneman1993,spitkovsky2005} to simulate collisionless shocks.
We send a nonrelativistic, supersonic, super-Alfv\'enic, electron-proton plasma flow with velocity $v_\text{u}$ against a reflecting wall placed at $x=0$; 
the interaction between incoming and reflected flows produces a shock moving along $x$ (to the right in figures); 
therefore, the shock is modeled in the downstream frame.
The computational box is 1D along $x$, with all components of fields and velocities retained. 
In order to save computational resources, the box is enlarged to $\sim 4\times10^5$ cells by expanding the right  boundary as simulation proceeds. 
The resolution is 10 cells per electron skin depth $c/\omega_\text{pe}$, where $c$ is the speed of light and $\omega_\text{pe}=\sqrt{4\pi n e^2/m_e}$ is the electron plasma frequency ($e$ and $n$ being the electric charge and number density); 
the time step is $\Delta t=0.045\omega_\text{pe}^{-1}$. 
We use 200 particles per cell per species, with a reduced proton-to-electron mass ratio $m_p/m_e=100$.
Electrons and protons are initially in thermal equilibrium, i.e., $T_e=T_p=1.12\times10^{-3}m_ec^2=1.12\times10^{-3}m_pv_u^2$, and $v_\text{u}=0.1c$.
The sonic and Alfv\'{e}nic Mach numbers are $M_s\equiv v_\text{sh}/\sqrt{T_p/m_p}=40$ and $M_A\equiv v_\text{sh}/v_A=20$, where $v_\text{sh}\equiv v_\text{u}r/(r-1)$ is the upstream flow speed in the shock rest frame and $v_A\equiv B_1/\sqrt{4\pi n m_p}$ is the Alfv\'{e}n speed in the initial magnetic field $\bold{B}_1=B_{1}(\cos\vartheta{\bf x}+\sin\vartheta{\bf y})$, with $\vartheta=30^\circ$.
To improve performance, we also implemented dynamical load balancing that repartitions the domain and particles amongst CPUs to even out the load. 

\begin{figure}
\begin{center}
\includegraphics[width=0.49\textwidth] {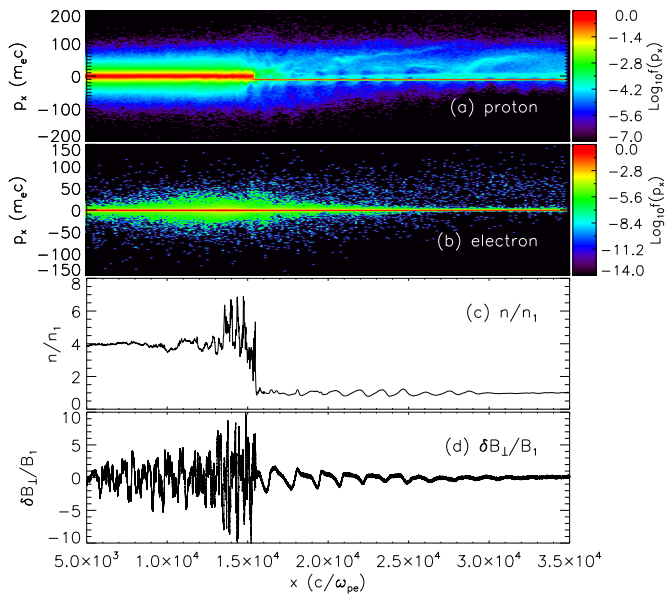}
\caption{(color online) Proton (a) and electron (b) $x-p_x$ phase space distributions for $f(p_x)$, density profile (c), and transverse magnetic field (d) at $t=4.6\times10^5\omega_\text{pe}^{-1}$ for a shock with $v_{\rm u}=0.1c$, $\vartheta=30^\circ$, and $m_p/m_e=100$.
Energetic protons and electrons diffuse ahead of the shock, amplifying the upstream magnetic field.}
\label{fig:shockstructure}
\end{center}
\end{figure}

\emph{Bell Instability.---} Fig.~\ref{fig:shockstructure} shows the proton (a) and the electron (b) $x-p_x$ phase space distribution for $f(p_x)$ at the end of our simulation at $t\simeq 4.6\times10^5\omega_\text{pe}^{-1}\approx 310\Omega_\text{cp}^{-1}$, where $\Omega_\text{cp}\equiv eB_1/{m_pc}$ is the proton cyclotron frequency. The streaming energetic protons and electrons are prominent in the upstream region ($x > 1.55\times10^4 c/\omega_\text{pe}$). The density is compressed by the expected factor of $4$ at the shock (Fig.~\ref{fig:shockstructure}c). 
The super-Alfv\'enic streaming of energetic protons excites magnetic turbulence in the upstream via the fast non-resonant (Bell) instability \cite[][]{bell2004}.
Fig.~\ref{fig:shockstructure}(d) shows the self-generated magnetic field, i.e., the field component $\delta B_\perp$ transverse to $\bold{B}_1$: 
the magnetic field is amplified by a factor of $\sim 2$ in a region of width $\sim 5\times10^3c/\omega_\text{pe}$ upstream of the shock (shock precursor), in agreement with the saturation level $\delta B_\perp/B_1\sim M_A\sqrt{\eta v_\text{sh}/c}$ of the Bell instability \cite[e.g.,][]{amato2009}.

\begin{figure}
\begin{center}
\includegraphics[width=8.6cm] {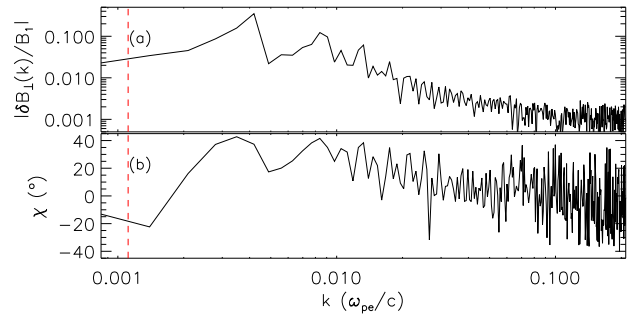}
\caption{(color online) (a) Spectral distribution of the perpendicular magnetic field, $\delta{B_\perp}/B_1$, in the shock precursor.
(b) Polarization angle $\chi(k)$, where $\chi=+(-)45^\circ$ corresponds to right-(left-)handed circularly polarized modes. 
The red dashed line indicates the inverse of the mean CR gyro-radius.}
\label{fig:fieldspectrum}
\end{center}
\end{figure}

To assess the nature of the excited modes, we performed Fourier analysis of $\delta B_\perp$ in the precursor (Fig.~\ref{fig:fieldspectrum}a), finding that the wave spectral energy density peaks at wavenumber $k_\text{max}=4\times10^{-3}\omega_\text{pe}/c>k_{\rm res}$, where $k_{\rm res}\equiv 1/\rho_\text{CR}\simeq 1.1\times10^{-3}\omega_\text{pe}/c$  is the wavenumber resonant with protons that contribute to the CR current (red dashed line in Fig.~\ref{fig:fieldspectrum}a).
More precisely, the relevant CR gyroradius, $\rho_\text{CR}$, is calculated by averaging over the distribution of non-thermal protons in the far upstream ($x>3\times10^{4}c/\omega_\text{pe}$), where the Bell instability is triggered \citep{caprioli2014b}.
Fig.~\ref{fig:fieldspectrum}(b) shows the polarization angle $\chi\equiv\sin^{-1}(V/I)/2$, where $I$ and $V$ are the Stokes parameters \citep{rybicki1979} for the two transverse magnetic field components in $k$-space.
Since $\chi=+(-)45^\circ$ corresponds to a right-(left-)handed circularly polarized waves, we conclude that modes with $k=k_\text{max}$ are indeed non-resonant Bell modes, while the mode at $k=1/\rho_{CR}$ is the resonant left-handed proton-cyclotron mode.
We note that short-wavelength, right-handed Bell modes are ineffective at disrupting the proton current, which allows the generated turbulence to grow to nonlinear levels \citep{caprioli2014b}, but very effective at scattering electrons, which can easily meet the cyclotron resonance criterion $k_{\max}\rho_e\sim 1$.

\begin{figure}
\begin{center}
\includegraphics[trim=4px 0px 0px 0px, clip=true,width=0.50\textwidth] {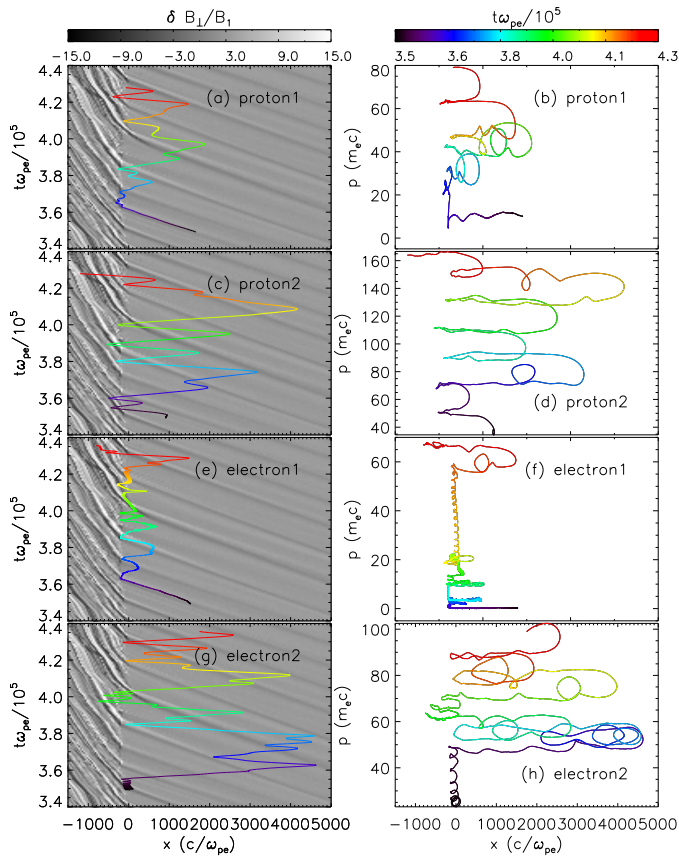}
\caption{(color online) 
Trajectories of individual protons (a-d) and electrons (e-h) in the $x-t$ and $x-p$ spaces (left and right columns).
The gray-scale colormap and the color code indicate the amplified magnetic field $\delta B_\perp/B_1$ and time, as in the legends.}
\label{fig:tracking}
\end{center}
\end{figure}

\emph{Proton and Electron Acceleration.---} 
In order to illustrate how protons and electrons achieve non-thermal energies, we tracked individual particles along their space-time trajectories.
In Fig. \ref{fig:tracking}, we follow two protons (a-d) and two electrons (e-h) over the time interval $3.4\times10^5<\omega_\text{pe} t <4.4\times10^5$. 
Left panels show particle trajectories in the $x-t$ plane (color lines), on top of the map of amplified magnetic field, $\delta B_\perp/B_1$ (grey scale), while right panels illustrate the momentum evolution along the particle trajectories;
positions are in the shock rest frame, and $p$ is calculated in the simulation frame. Panels (a-b) depict a proton with initial momentum $p=10m_ec(=0.1m_pc=m_pv_u)$ that encounters the shock at $t\approx 3.62\times10^5\omega_\text{pe}^{-1}$, gains energy in a few gyro-cycles of SDA around $t\approx 3.7\times10^5\omega_\text{pe}^{-1}$, and finally enters DSA at $t\approx 3.9\times10^5\omega_\text{pe}^{-1}$; when the proton is injected into DSA, its momentum is $p_\text{inj}\approx 3m_p v_{\rm u}=30 m_ec$, consistent with the model of injection from hybrid simulations in Ref.~\citep{caprioli2014d}.
In Fig.~\ref{fig:tracking}(c-d), a proton which has already been injected is shown to cross the shock several times and gains energy by undergoing head-on collisions against upstream waves (since the simulation frame is the downstream frame, post-shock reflections do not increase particle energy); its diffusion length, i.e., the maximum displacement from the shock, is now larger than its gyroradius.

\begin{figure}
\begin{center}
\includegraphics[trim=8px 0px 35px 0px, clip=true,width=0.49\textwidth] {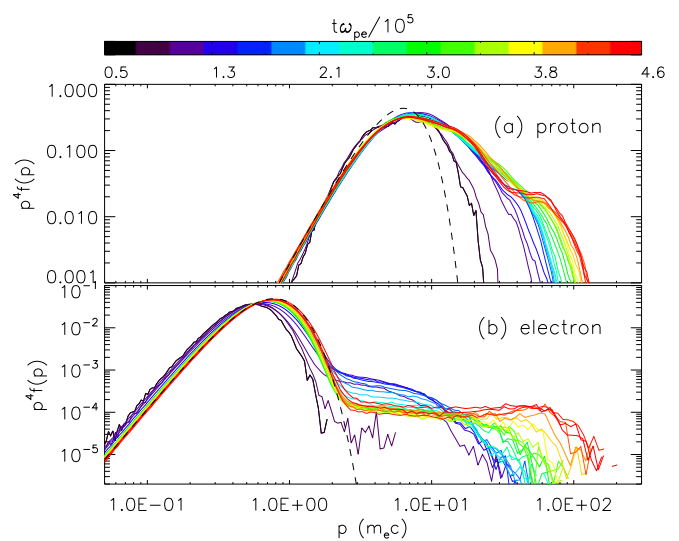}
\caption{(color online) Evolution of the downstream momentum distributions for (a) protons and (b) electrons.
The dashed lines represent thermal Maxwellian distributions.}
\label{fig:spectrum_evolution_m100}
\end{center}
\end{figure}

Electron acceleration proceeds in a different way.
In Fig.~\ref{fig:tracking}(e-f) we follow an initially cold electron that, after being reflected off the shock at $t\approx 3.65\times10^5\omega_\text{pe}^{-1}$ because of magnetic mirroring \cite[e.g.,][]{ball2001,park2013}, remains trapped between the shock front and the nearest upstream wave until $t\approx 4.25\times10^5\omega_\text{pe}^{-1}$.
At each interaction with the shock, the electron may undergo a new cycle of SDA, which results in vigorous energy gain.
This \emph{hybrid} (SDA+scattering on Bell waves) acceleration process is rather fast: in less than $8\times10^4\omega_\text{pe}^{-1}\simeq 50\Omega_{\rm cp}^{-1}$ electrons increased their energy by a factor of more than $10^4$ (from $5\times 10^{-3}m_ec^2$ to $60m_ec^2$). 
Guo \emph{et al.} studied the physics of a similar process for quasiperpendicular shocks \cite[\S4.2.3 in Ref.~][]{guo2014a}, in which the upstream confinement is provided by electron-induced firehose modes \cite{guo2014b} rather than by proton-induced Bell modes. 
In our longer simulations, we find that this hybrid acceleration transitions to standard DSA when the electron achieves a momentum $p_{\rm inj}\sim50m_ec$ (Fig.~\ref{fig:tracking}g,h).
Note that in this stage electrons diffuse into the upstream for more than one gyro-radius, gaining energy when they reverse their motion and not at the shock transition.
From other electron trajectories (not shown here), we observe that the typical momentum needed for injection into DSA spans the range $p\approx 30-100 m_ec$, comparable with the typical proton injection momentum $p\approx 3m_p  v_{\rm u}=30m_ec$ \cite{caprioli2014d}.

\emph{Proton and Electron Spectra.---} Fig.~\ref{fig:spectrum_evolution_m100} shows the time evolution of the proton and electron momentum distributions in a region $2000c/\omega_\text{pe}$ behind the shock, multiplied by $p^4$ to emphasize the scaling with the expected universal DSA spectrum at strong shocks.
The dashed lines represent the fitting with thermal (Maxwellian) distributions with almost the same temperature  for both electrons and protons ($T\approx 0.12m_ec^2=0.12m_pv_u^2$), attesting to very effective thermal equilibration between the two species.
The maximum energy of each species increases with time, and both species develop power-law distributions $\propto p^{-4}$ after $t\approx 4\times10^5\omega_\text{pe}^{-1}\simeq 268\Omega_\text{cp}^{-1}$, in remarkable agreement with the DSA prediction.
Very interestingly, electrons show the typical DSA slope even in the range of momenta where they undergo hybrid acceleration ($2.5\lesssim p/m_ec\lesssim30$), which is likely a manifestation of the fact that the balance between energy gain and escape probability per cycle of hybrid acceleration is more similar to DSA than to SDA \cite{bell1978}.

\emph{Electron/Proton Acceleration Efficiency.---} A very important quantity that can be measured in our simulations is the non-thermal electron-to-proton ratio, defined by
$K_\text{ep}\equiv f_{\rm e}(p)/f_{\rm p}(p)$, for $p\gtrsim p_{\rm inj}$. 
Since electron and proton CR distributions have the same spectral index, $K_{\rm ep}$ is independent of $p$.
From Fig.~\ref{fig:spectrum_evolution_m100} we infer $K_\text{ep}\approx 3.8\times10^{-3}$ for our reference case with $m_p/m_e=100$ and $v_\text{u}/c=0.1$.

Since $K_\text{ep}$ may be affected by the reduced proton-to-electron mass ratio, we performed additional simulations to investigate such a dependence.
With all the other shock parameters left unchanged, we find $K_\text{ep}\approx 5.5\times10^{-3}$ for $m_p/m_e=400$; 
we interpret such a marginal increase from $m_p/m_e=100$ to $m_p/m_e=400$ as due to post-shock thermal electrons becoming trans-relativistic when sharing the proton temperature.
Extrapolating such a trend to the realistic mass ratio would lead to $K_{\rm ep}\lesssim 0.01$.
Such high shock velocity is relevant for shocks in radio-SNe \cite[e.g.,][]{cf06}, i.e., radio-bright extragalactic SNe observed a few days to a year after  explosion.

We have also investigated the case of lower shock velocities relevant for young Galactic SNRs (such as Tycho, Cas A, SN1006, Kepler, etc.), whose blast waves travel at $v_\text{u}\sim 0.01-0.02c$.
For $v_\text{u}/c=0.05$ and $m_p/m_e=100$ we find $K_\text{ep}=1.2\times10^{-3}$, a value a factor of $\sim 3$ smaller than for $v_\text{u}/c=0.1$, which implies that $K_{\rm ep}$ is roughly proportional to $v_\text{sh}/c$.
The extrapolation to the case of young SNRs in both mass ratio and shock velocity suggests that the CR electron/proton ratio should be in the range $K_{\rm ep}\approx 1-3\times 10^{-3}$, in good agreement with the value $K_\text{ep}\approx 1.6\times10^{-3}$ inferred in  Tycho's SNR \citep{morlino2012}, and only a factor of few different from the $K_{\rm ep}$ measured at Earth around 10 GeV.
Quite interestingly, our results suggest that such a ratio in Galactic CRs may be accounted for if the bulk of CRs are accelerated during the early stages of the SNR evolution, when the shock is quite fast. 
Nevertheless, even if the quoted values of $K_{\rm ep}$ seem not to vary at later times in our simulations, an extended analysis of the space of the shock parameters ($M_A$, $\beta$, $\vartheta$, in addition to $m_p/m_e$ and $v_\text{sh}$) is needed to cover the vast non-thermal phenomenology of radio SNe and SNRs.

\emph{Conclusions.---} We studied proton and electron acceleration in non-relativistic high Mach number $(M_A=20)$ quasiparallel $(\vartheta=30^\circ)$ shocks using 1D PIC simulations, attesting for the first time the simultaneous DSA of both species, which leads to the development of the universal power-law momentum distributions $\propto p^{-4}$. 
Protons are efficiently accelerated at quasiparallel shocks \cite{caprioli2014a} and for high $M_A$ excite the non-resonant (Bell) streaming instability \cite{bell2004,caprioli2014b}. 
Strong Bell modes enable electron acceleration by making the shock locally oblique, which allows a large pitch angle for mirror reflection,
and providing small-wavelength, right-handed modes that enhance electron scattering and facilitate their rapid return to the shock. 
Protons are injected into DSA after a few gyro-cycles of SDA, while electrons are first pre-heated via SDA, and then enter a hybrid stage in which they are scattered by the upstream Bell waves and rapidly sent back to the shock for more SDA cycles. They enter DSA when their momentum is comparable to the proton injection momentum, $p_{\rm inj}\sim 3m_p v_{\rm u}$.

From our simulations, the CR electron-to-proton ratio is inferred to be $K_\text{ep}\approx 10^{-3}-10^{-2}$ for  $v_\text{u}/c\approx 0.02-0.1$, close to the values observed in young Galactic SNRs \citep[e.g.,][]{morlino2012} and in the CR fluxes measured at Earth. The lower electron acceleration efficiency is due to the additional cycles of hybrid acceleration that electrons have to undergo to reach the injection momentum, and the probability of loss at each shock encounter. 

Our simulations provide a self-consistent interpretation of the simultaneous acceleration of protons and electrons as a function of magnetic inclination. 
Hybrid simulations \cite{caprioli2014a} show that proton DSA arises naturally for quasiparallel shocks, while quasiperpendicular shocks do not reflect protons into the upstream in the absence of strong pre-existing turbulence. 
Such a problem with proton \emph{injection} does not
preclude rapid acceleration of previously pre-energized protons even at quasiperpendicular shocks \citep[e.g.,][]{Ferrand+14}. 
Shocks of all inclinations seem to reflect electrons into the upstream; whether these electrons return to participate in DSA depends on the presence of upstream turbulence. 

We show that proton-driven waves in quasiparallel shocks are effective at trapping electrons. It has been recently shown \cite{guo2014b} that, for quasiperpendicular shocks propagating in high-$\beta$ plasmas ($\beta\equiv 2M_A^2/M_s^2$), pre-accelerated electrons can excite low-amplitude upstream waves, which may act as scattering agents and cause electron  diffusion. 
At the moment, no PIC simulations of initially quasiperpendicular shocks show conclusive evidence of electron DSA, although we expect this is just a matter of insufficient simulation size. 
We conjecture that quasiperpendicular shocks should be able to generate electron-driven upstream turbulence that will eventually lead to electron DSA. 
As a function of magnetic inclination, then, quasiparallel shocks should be efficient at both the electron and proton acceleration, while quasiperpendicular shocks should show electron acceleration, with virtually no proton acceleration. Future simulations will allow us to quantify these relative efficiencies for comparison with SNR observations, where obliquity of shocks can be measured with radio polarization.

\begin{acknowledgments}
This research was supported by NASA (grant NNX14AQ34G to DC) and Simons Foundation (grants 267233 and 291817 to AS), and facilitated by Max-Planck/Princeton Center for Plasma Physics. Simulations were performed on computational resources provided by Princeton High-Performance Computing Center, by NERSC (supported by the Office of Science of the U.S. Department of Energy under Contract No. DE-AC02-05CH11231), and by XSEDE (allocation No. TG-AST100035).
\end{acknowledgments}

\end{document}